\documentclass[12pt]{iopart}
\usepackage{iopams}
\usepackage{graphicx}
\usepackage{siunitx}
\usepackage{hyperref}
\usepackage{microtype}
\usepackage{xcolor}
\hypersetup{colorlinks=true, linkcolor=blue, citecolor=blue, urlcolor=blue}
\usepackage[switch]{lineno}

\begin{document}

\title[Temperature-dependent interlayer exchange in SAF]{Thermally-controlled interlayer exchange and field-induced anisotropy in synthetic antiferromagnets}

\author{O~Kozlov$^{1,2}$, V~Kalita$^{1,2,3}$, S~Reshetniak$^{1,2}$, A~Kravets$^{2,4}$, D~Polishchuk$^{2,4}$, and V~Korenivski$^4$  }

\address{$^1$ National Technical University of Ukraine ``Igor Sikorsky Kyiv Polytechnic Institute'', 03056 Kyiv, Ukraine}
\address{$^2$ V.G. Baryakhtar Institute of Magnetism, NAS of Ukraine, 03142 Kyiv, Ukraine}
\address{$^3$ Institute of Physics, NAS of Ukraine, 03028 Kyiv, Ukraine}
\address{$^4$ Nanostructure Physics, Royal Institute of Technology, 10691 Stockholm, Sweden}

\ead{anatolii@kth.se}

\begin{abstract}

Interlayer exchange in synthetic antiferromagnets incorporating thin paramagnetic spacers can be controlled thermally. The spacer provides an additional ferromagnetic contribution that renormalizes the otherwise temperature-independent interlayer coupling. As a result, the system shows antiferromagnetic alignment at high temperatures and ferromagnetic alignment at low temperatures. This behavior is observed in Fe(2 nm)/Cr(0.4 nm)/Fe$_{17.5}$Cr$_{82.5}$(0.9 nm)/Cr(0.4 nm)/Fe(2 nm) multilayers with the inner spacer Fe$_{17.5}$Cr$_{82.5}$ paramagnetic at and above room temperature, and is shown to be due to the spacer being significantly magnetically polarized on lowering the temperature toward its Curie point. Although the Fe layers lack intrinsic magnetocrystalline anisotropy, the magnetization reversal demonstrates a field-induced uniaxial anisotropy of antiferromagnetic character. The resulting reversal process resembles that of a metamagnet with a spin-flip transition.

\end{abstract}


\section{Introduction}

Magnetic multilayers where ferromagnetic (FM) layers are coupled by the Ruderman–Kittel–Kasuya–Yosida (RKKY) exchange interaction \cite{Bruno1991} remain a subject of intense research due to their unique magnetic properties and many technological applications \cite{Fert1995,Grunberg2000,Grunberg2001}. In a typical such system, nonmagnetic (NM) spacers separating the FM layers mediate an oscillatory and rapidly decaying interlayer exchange interaction with a characteristic length of $\approx$~1--4 nm \cite{Parkin1990,Parkin1991,Egelhoff1992}. Depending on the spacer thickness, the sign of the coupling alternates, giving rise either to ferromagnetic (FM) or antiferromagnetic (AFM) alignment of the FM-layer magnetizations. When the AFM configuration is realized in a FM/NM/FM trilayer, the structure forms a synthetic antiferromagnet (SAF), which is a key functional element in many modern spintronic devices \cite{Duine2018}. SAFs are widely used in magnetoresistive sensors \cite{Berg1996,Leal1998,Zhu1999,Veloso2000,Parkin2003}, magnetic random-access memory (MRAM) \cite{Apalkov2016,Bhatti2017,Parkin2003,Bergman2011}, spin-torque nano-oscillators \cite{Houssameddine2010,Firastrau2013,Jiang2019,Zhou2020}, and spin-logic devices \cite{Slaughter2002,Dieny2020,Fattouhi2021}. Compared to single FM layers, SAFs provide enhanced thermal stability, reduced stray fields, faster switching dynamics \cite{Yang2015,Kravets_2015,Kravets_2016,Dzhezherya2025,Guo2025}, and lower critical currents for spin-torque-driven magnetization reversal.

Within the RKKY mechanism, the magnetic moment of one FM layer polarizes the conduction electrons in the NM spacer, generating oscillations in spin density that mediate the interaction with the second FM layer \cite{Bruno1991,Bruno1995}. For a simple model, the exchange energy $J(r)$ is proportional to $\cos(2 k_{F}r)$, where $r$ is the distance between magnetic moments and $k_{F}$ is the Fermi wave vector. The sign $\cos(2 k_{F}r)$ determines whether the interaction is FM or AFM.

A prototypical example of SAF is the three-layer structure Fe/Cr/Fe \cite{Grunberg1986}, where strong AFM coupling occurs for a Cr thickness of $\sim 1$~nm with a period of oscillations of $\sim 2$~nm \cite{Parkin1990}. These structures led to the discovery of giant magnetoresistance (GMR) \cite{Baibich1988,Grunberg1989} and laid the foundation for spintronics \cite{Zutik2004}. Oscillatory RKKY coupling has been observed across a broad range of multilayers composed of 3d, 4d, and 5d transition metals \cite{Parkin1991,Egelhoff1992}.

Because the RKKY coupling is governed by electrons near the Fermi level, thermal smearing of the Fermi–Dirac distribution has only a weak effect at room temperature. In Fe/Cr/Fe, the AFM coupling remains robust up to 400–500 K \cite{Fullerton1995,Drovosekov2001}. The most significant factors affecting the coupling strength are the interface roughness and spacer-thickness variations \cite{Pierce1999}. Temperature effects become more pronounced when the spacer thickness is increased and the RKKY coupling weakens.

A qualitatively different behavior arises when the NM spacer contains localized magnetic moments, such as Fe impurities in Cr \cite{Polishchuk2017epl,Polishchuk2017prb}, Ni impurities in Cu \cite{Parkin1993}, or Fe and Co impurities in Ru \cite{Winther2024}, forming a dilute ferromagnetic (f) layer in FM/f/FM. At low impurity concentrations, the f-layer remains predominantly paramagnetic and continues to mediate RKKY coupling, albeit with additional spin scattering \cite{Parkin1993,Polishchuk2017epl,Polishchuk2017prb}. At higher concentrations, however, the f-layer may develop collective ferromagnetic order below a certain temperature, which disrupts the classical oscillatory RKKY behavior and mixes the RKKY exchange with a direct exchange component mediated by the magnetized spacer. As a result, the effective interlayer FM–FM coupling becomes strongly temperature dependent \cite{Polishchuk2017prb}.

This interplay leads to temperature-driven switching between magnetic states: at low temperatures the structure behaves as a synthetic ferromagnet due to the direct exchange through the magnetized f-layer (combined with the strong FM-RKKY via the ultra-thin Cr at the two interfaces), while at elevated temperatures (with f paramagnetic) the system exhibits AFM RKKY coupling as in a conventional SAF \cite{Polishchuk2017prb}. Such behavior has been demonstrated in  Fe(2~nm)/Cr(0.4~nm)/Fe$_{17.5}$Cr$_{82.5}$(0.9~nm)/Cr(0.4~nm)/Fe(2~nm) when heated by an electric current \cite{Iurchuk2023}. 

Because the AFM RKKY exchange energy is proportional to the scalar product of the FM-layer magnetizations \cite{Fert1995}, SAFs share many features with antiferromagnets when subjected to external magnetic fields and may exhibit field-induced criticality as to transitions between the AFM and FM states. Experimental studies reveal step-like changes in magnetoresistance indicative of abrupt reorientation events \cite{Iurchuk2023}. In contrast to classical antiferromagnets, however, the magnetization reversal in SAFs often displays hysteresis, which is almost independent of the in-plane field direction, contradicting models based on uniaxial anisotropy. This discrepancy has been reported for Fe/Cr/FeCr/Cr/Fe \cite{Polishchuk2018nrl}.

In this work, we investigate the magnetization reversal of the five-layer structure Fe(2~nm)/Cr(0.4~nm)/Fe$_{17.5}$Cr$_{82.5}$(0.9~nm)/Cr(0.4~nm)/Fe(2~nm), characterized by a temperature-dependent interlayer exchange, with the Fe$_{17.5}$Cr$_{82.5}$ spacer being paramagnetic at room temperature. We propose a method to determine the critical magnetic fields associated with the abrupt transformations of the magnetic state. We show that these critical fields result from the combined action of two interlayer exchange contributions -- an AFM component that is essentially temperature independent and a FM component that follows the temperature dependence of the Fe$_{17.5}$Cr$_{82.5}$ susceptibility.

To account for the experimentally observed isotropy of the in-plane magnetization, we employ a model of rotational magnetic anisotropy, in which the easy axis aligns with the external magnetic field after saturation. The magnitude of this anisotropy field is comparable to the interlayer exchange fields, leading to a magnetization behavior reminiscent of metamagnetic transitions.

\section{Samples}

\begin{figure}
\centering
\includegraphics[width=0.4\linewidth]{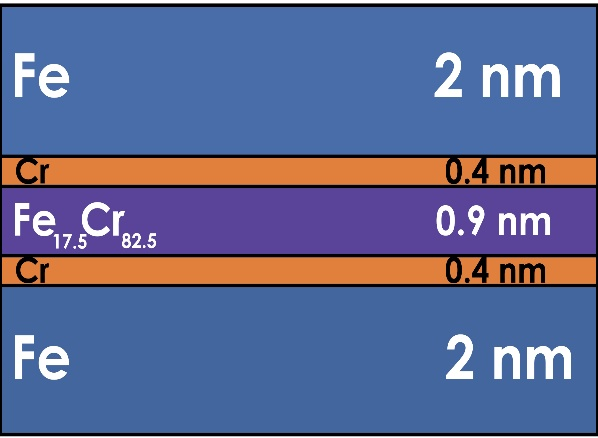}
\caption{Schematic of studied multilayer: two ferromagnetic Fe layers separated by two nonmagnetic Cr spacers and central Fe$_{17.5}$Cr$_{82.5}$ layer.}
\label{figure1}
\end{figure}

Multilayers of Fe(2~nm) / Cr(0.4~nm) / Fe$_{17.5}$Cr$_{82.5}$(0.9~nm) / Cr(0.4~nm) / Fe(2~nm) (Fig/~\ref{figure1}) were deposited by dc magnetron sputtering (AJA International) onto Si(100) substrates pretreated by Ar ion etching. The Fe$_{17.5}$Cr$_{82.5}$ layer was grown by co-sputtering from separate Fe and Cr targets. 

Fe$_x$Cr$_{1-x}$ alloys are dilute ferromagnets whose Curie temperature depends on the Fe concentration $x$ \cite{Babic1980,Burke1978}. Such alloys have previously been incorporated in Fe/Fe$_x$Cr$_{1-x}$/Fe trilayers, where the temperature-driven transition of the Fe$_x$Cr$_{1-x}$ spacer from ferromagnetic to paramagnetic leads to a switch in the Fe–Fe interlayer coupling from ferromagnetic to antiferromagnetic \cite{Polishchuk2017epl,Polishchuk2017prb}.

The magnetic moment was measured using a MPMS3 SQUID magnetometer in the temperature range 200–400~K. The field was applied parallel to the film plane and the detected signal corresponded to the component of the magnetic moment along the applied field $\mathbf{H}$.

Insertion of additional Cr spacers between Fe and Fe$_x$Cr$_{1-x}$ has been shown to improve the thermal switching of the interlayer coupling \cite{Polishchuk2017epl,Polishchuk2017prb}, and current-induced heating can be used to control the Fe layer switching fields \cite{Iurchuk2023}.

The Curie temperature of the 0.9~nm Fe$_{17.5}$Cr$_{82.5}$ layer is below 200~K. Thus, in the studied temperature range the central layer remains paramagnetic, while the Fe layers retain ferromagnetic order.

\section{Results}

\subsection{Hysteresis loops}

\begin{figure}[t]
\centering
\includegraphics[width=0.6\linewidth]{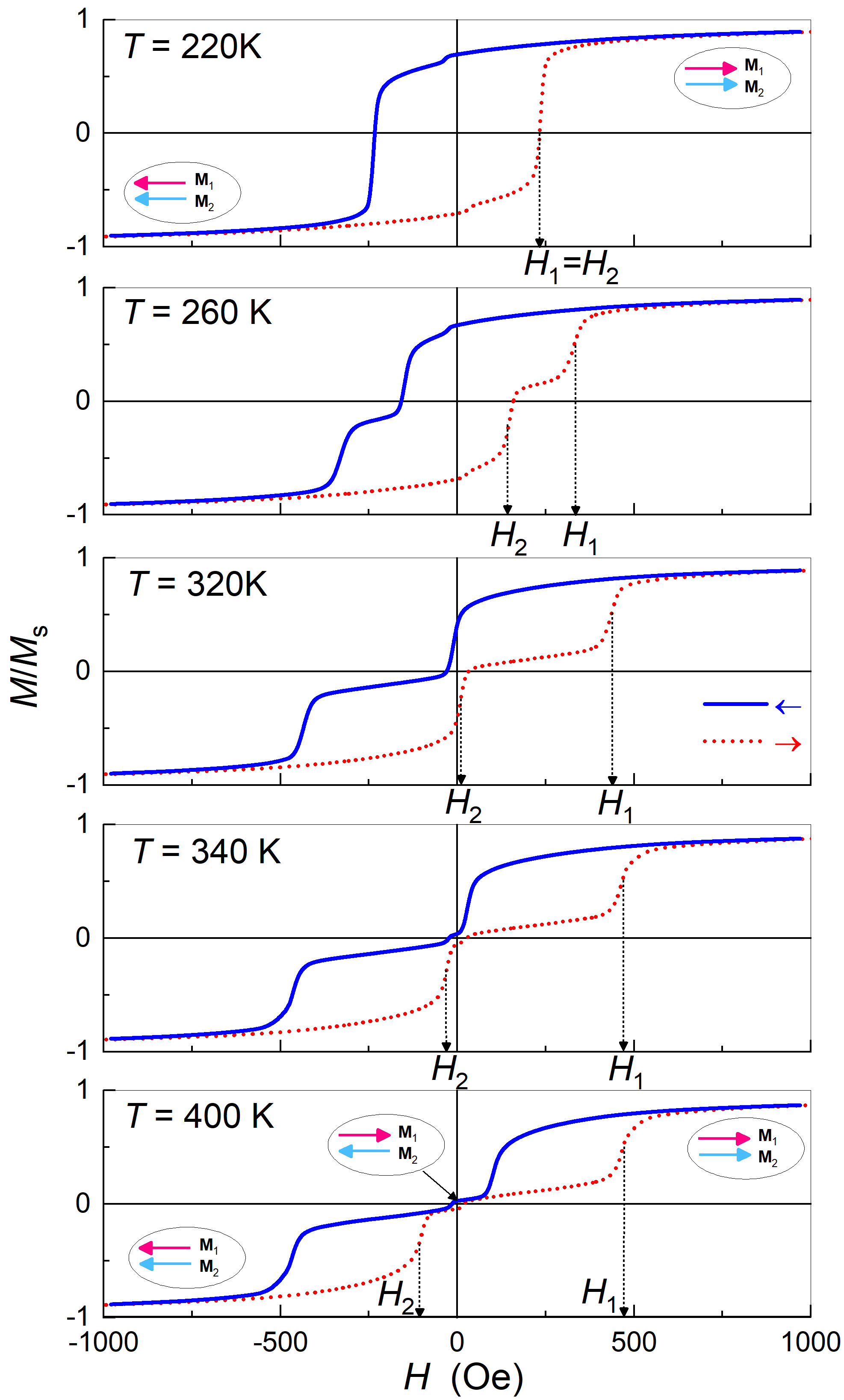}
\caption{Field dependence of normalized magnetization $M/M_{\text{s}}$ at various temperatures.}
\label{figure2}
\end{figure}

Figure~\ref{figure2} shows the field dependence of the normalized magnetization, $M/M_{\text{s}}$, where $M_{\text{s}}$ is the saturation magnetization and $M$ is its projection along the applied field $\mathbf{H}$. The curves exhibit typical hysteresis loop shapes. Arrows indicate the field-sweep direction: leftward for decreasing field and rightward for increasing field. Vector diagrams in Fig.~\ref{figure2} illustrate the magnetization orientations in the ferromagnetic layers during reversal.

At $T=220$~K, the loops are characteristic of a uniaxial ferromagnet with the field applied along the easy axis. Magnetization reversal occurs by a simultaneous switch of the magnetizations in both Fe layers.

With increasing temperature, the reversal mechanism changes: the layers no longer switch simultaneously, and the loops evolve into a superposition of two shifted minor loops, one for each layer. The shift increases with temperature. In Fig.~\ref{figure2}, the decreasing-field branch is shown in blue and the increasing-field branch is shown in red. At $T=340$~K and $400$~K, the net magnetization at $H=0$ vanishes, consistent with antiferromagnetic alignment.

Antiferromagnetic alignment is also present at lower temperatures, but the loops retain a finite remanence at $H \neq 0$. Furthermore, when the external field is applied in-plane at arbitrary orientations, the loops remain unchanged. This isotropy indicates the absence of intrinsic magnetocrystalline anisotropy and suggests the presence of a field-induced uniaxial anisotropy, with the easy axis aligned along the applied field.

\subsection{Critical fields}

\begin{figure}[t]
\centering
\includegraphics[width=0.6\linewidth]{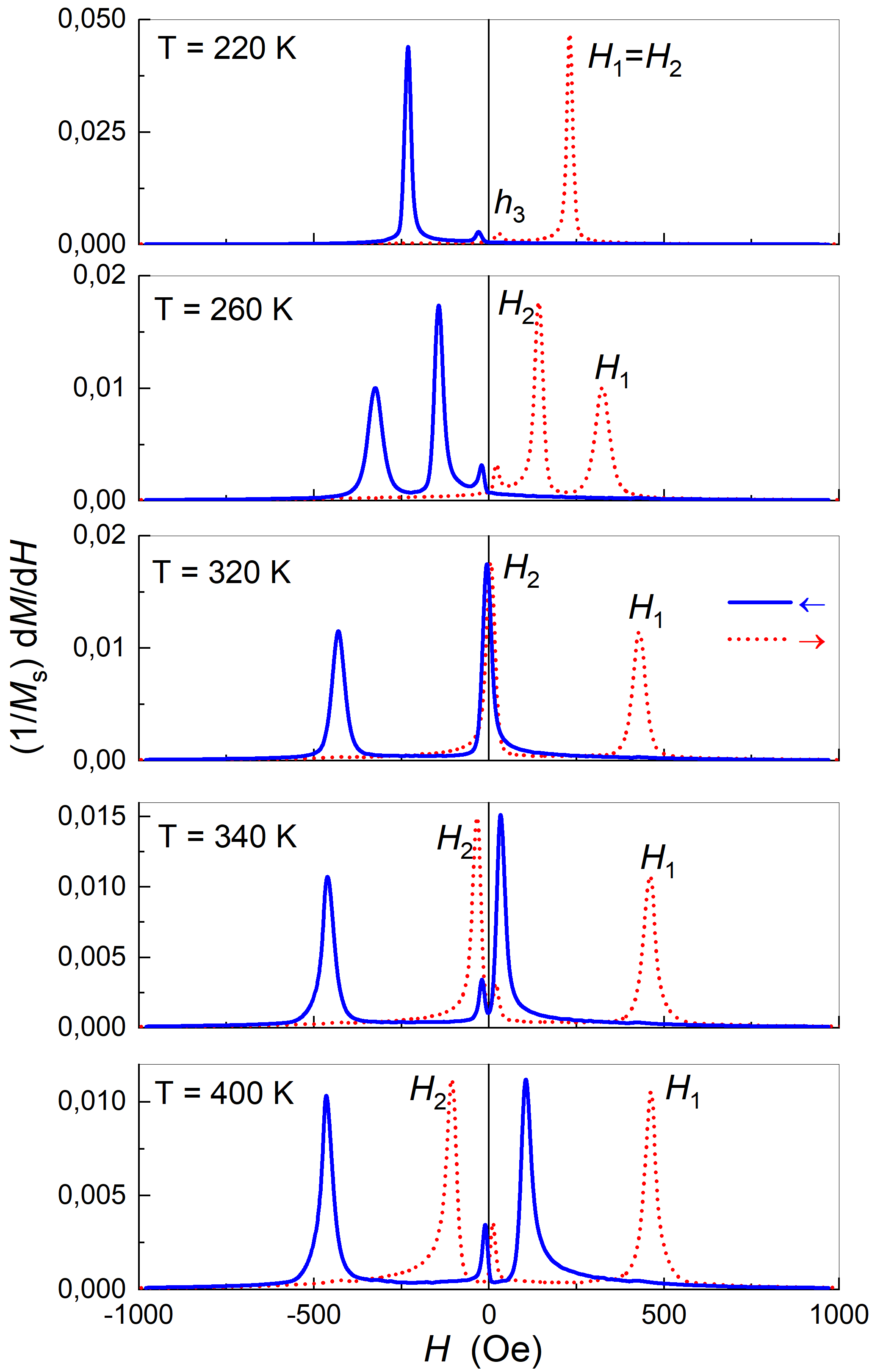}
\caption{Field dependence of $(1/M_{\text{s}})(dM/dH)$ at various temperatures, calculated from magnetization curves in Fig.~\ref{figure2}.}
\label{figure3}
\end{figure}

Figure~\ref{figure3} shows the field dependence of the derivative of the normalized magnetization, $(1/M_{\text{s}})(dM/dH)$, obtained from the curves in Fig.~\ref{figure2}. The derivative exhibits pronounced peaks corresponding to the critical fields at which the magnetic state of the film changes.

At $T=200$~K, when the film displays ferromagnetic behavior, the peaks coincide with the abrupt changes in the magnetization occurring when the applied field reaches the effective anisotropy field. This behavior is characteristic of a uniaxial ferromagnet.

At higher temperatures, the magnetization reversal acquires features of the antiferromagnetic interlayer ordering, evidenced by the doubling of maxima in the derivative curves. In Fig.~\ref{figure3}, this appears as two distinct peaks on each branch of the hysteresis loop. This behavior arises from the interlayer exchange interactions.

\begin{figure}[t]
\centering
\includegraphics[width=0.6\linewidth]{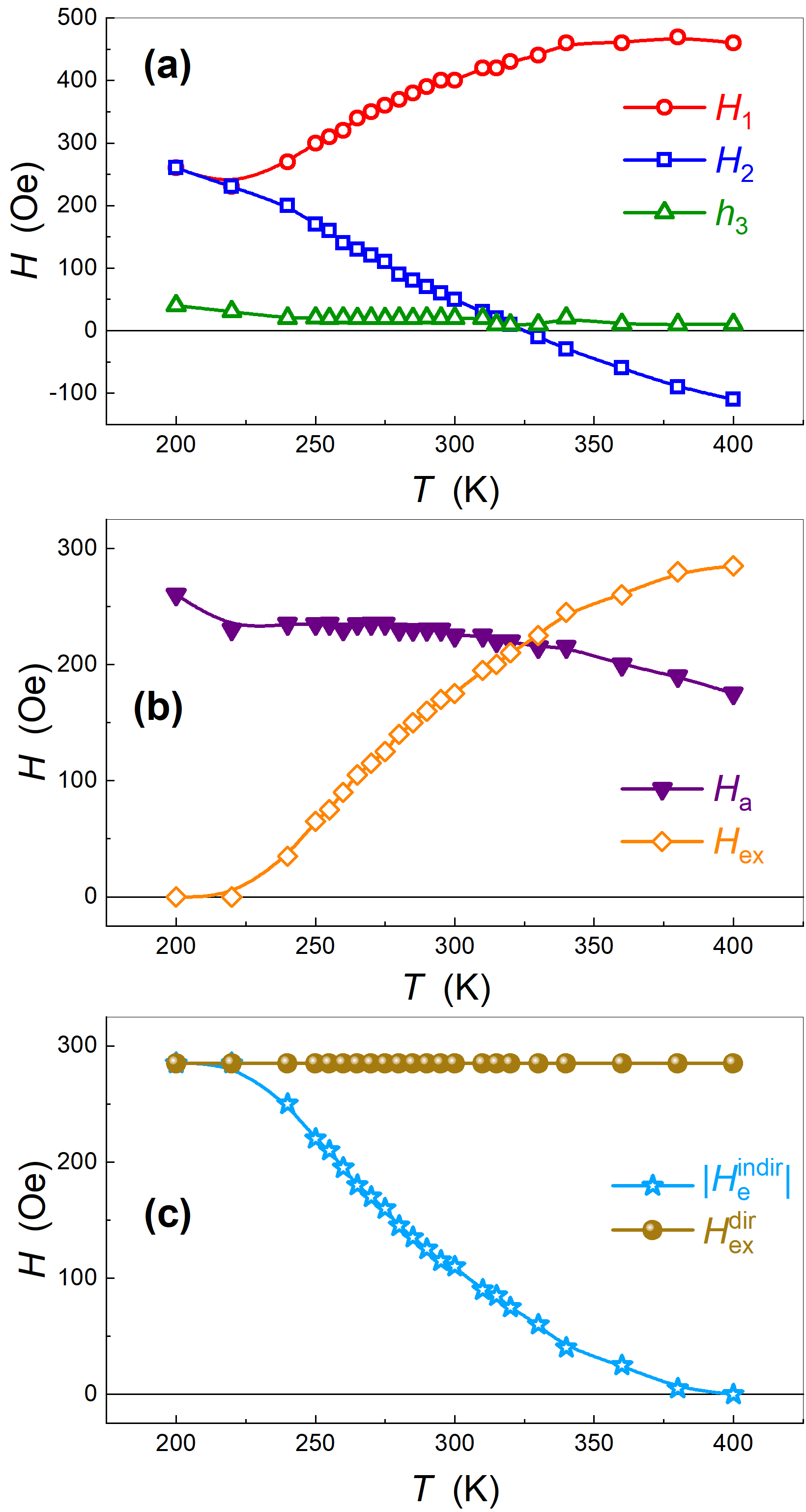}
\caption{(a) Temperature dependence of $H_1$ and $H_2$; (b) anisotropy field $H_{\text{a}}$ and effective exchange field $H_{\text{ex}}$; (c) direct and indirect RKKY-exchange fields $H_{\text{dir}}$ and $|H_{\text{indir}}|$.}
\label{figure4}
\end{figure}

The corresponding critical fields are denoted $H_1$ and $H_2$. Their temperature dependence, $H_1(T)$ and $H_2(T)$, are presented in Fig.~\ref{figure4}~(a).

Figure~\ref{figure4}~(a) also shows the temperature dependence of an additional field $h_3$, corresponding to the central peak in the derivative curves. Its value is considerably lower than $H_1$ and $H_2$ and exhibits only weak temperature dependence. This field is associated with a small ferromagnetic fraction in the film that does not participate in interlayer exchange.

\section{Discussion}

The critical behavior of the film's magnetization reversal and its temperature dependence can be explained by the combined effects of interlayer exchange and induced magnetic anisotropy.

\subsection{Interlayer Exchange}

The exchange energy density arising from the interaction between the ferromagnetically ordered Fe layers is proportional to the scalar product of their magnetizations, 
\begin{equation}
  E_1=J_{\text{dir}} \textbf{M}_1 \textbf{M}_2,    
\end{equation}
where $J_{\text{dir}}$ is the direct RKKY-exchange constant and $\mathbf{M}_1$, $\mathbf{M}_2$ are the layer magnetizations averaged over each layer.

In addition, the Fe$_{17.5}$Cr$_{82.5}$ spacer interacts with both Fe layers, giving an exchange contribution
\begin{equation}
    E_2 = J(\mathbf{M}_1 + \mathbf{M}_2)\mathbf{m},    
\end{equation}
where $J$ is the exchange constant between the spacer and Fe, and $\mathbf{m}$ is the magnetization of the spacer. Since the Fe$_{17.5}$Cr$_{82.5}$ layer is paramagnetic, $\mathbf{m}$ is induced by the exchange field
\begin{equation}
    \mathbf{h}_{\text{ex}} = -\frac{\partial E_2}{\partial \mathbf{m}} = -J(\mathbf{M}_1 + \mathbf{M}_2),     
\end{equation}
and can be written as $\mathbf{m} = \chi \mathbf{h}_{\text{ex}} = -\chi J(\mathbf{M}_1 + \mathbf{M}_2)$, with $\chi$ the magnetic susceptibility of the spacer.

For a thin spacer magnetized solely by the effective exchange field, the energy reduces to
\begin{equation}
    E_2 = -\frac{1}{2}\chi J^2 (\mathbf{M}_1 + \mathbf{M}_2)^2.
\end{equation}
Neglecting the constant terms $\mathbf{M}_1^2$ and $\mathbf{M}_2^2$, the relevant contribution is
\begin{equation}
  E_2 = -\chi J^2 \mathbf{M}_1 \mathbf{M}_2 .
\end{equation}
The total exchange energy of the film is then
\begin{equation}
     E_{\text{ex}} = E_1 + E_2 = \left( J_{\text{dir}} - \chi J^2 \right) \mathbf{M}_1 \mathbf{M}_2 ,
\label{eq:1}
\end{equation}
where the second term, $-\chi J^2$, represents the indirect, temperature-dependent RKKY-exchange mediated by the paramagnetic spacer. Since $J_{\text{indir}}<0$, this interaction is ferromagnetic. \emph{For brevity, in what follows, direct and indirect exchange refers to interlayer RKKY exchange}. 

For a spacer of finite thickness, the interlayer exchange energy takes the general form
\begin{equation}
      E_{\text{ex}} = \left( J_{\text{dir}} + J_{\text{indir}} \right) \mathbf{M}_1 \mathbf{M}_2 ,
\label{eq:2}
\end{equation}
with $J_{\text{indir}} < 0$ reflecting the paramagnetic response of the Fe$_{17.5}$Cr$_{82.5}$ layer.

Thus, when the direct RKKY-exchange between the Fe layers is antiferromagnetic, it competes with the ferromagnetic indirect RKKY-exchange. As temperature decreases, the indirect contribution strengthens due to the increase of the spacer susceptibility, leading to a crossover in the magnetization curves from antiferromagnetic-like behavior at high temperatures to ferromagnetic-like behavior at low temperatures, consistent with the experimental data in Fig.~\ref{figure2}.

\subsection{Induced magnetic anisotropy}

To describe the magnetization curves in Fig.~\ref{figure2}, it is necessary to consider not only the interlayer exchange but also magnetic anisotropy. Experimentally, the magnetization curves show no dependence on the in-plane orientation of the applied field, indicating the absence of intrinsic magnetocrystalline anisotropy with well-defined in-plane easy and hard axes. 
Nevertheless, the rectangular hysteresis loops resemble those of uniaxial magnets when the field is applied along the easy axis. This behavior can be attributed to induced magnetic anisotropy, which develops after the film is saturated by the external field \cite{Polishchuk2017epl}. The easy axis of this induced, rotating anisotropy aligns with the direction of the saturating field.

Taking the $z$ axis along the applied field ($\mathbf{H} \parallel \mathbf{z}$), the uniaxial anisotropy energy in the Fe layers can be written as
\begin{equation}
E_2 = -\frac{1}{2} K \left ( M^2_{1z} + M^2_{2z} \right ),
\label{eq:3}
\end{equation}
where $K$ is the field-induced anisotropy constant and $M_{1z}$ and $M_{2z}$ are the projections of the magnetizations $\mathbf{M}_1$ and $\mathbf{M}_2$ onto the $z$ axis.

This anisotropy acts only as an orienting factor for the magnetization vectors, while their magnitudes remain fixed at the saturation value, $|M_{1z}| = |M_{2z}| = M_{\text{s}}$.

\subsection{Temperature dependence of interlayer exchange}

The total energy $E$ of the film’s magnetic configurations in an external field $\mathbf{H} \neq 0$, with magnetization vectors as shown in Fig.~\ref{figure2}, can be written as
\begin{eqnarray}
    E&=& E_{\text{ex}} + E_a - \mathbf{H} (\mathbf{M}_1 + \mathbf{M}_2)\nonumber\\
    &=& (J_{\text{dir}} + J_{\text{indir}}) M_{1z} M_{2z} 
    - \frac{1}{2} K(M_{1z}^2 + M_{2z}^2) - H_z (M_{1z} + M_{2z}) ,
\label{eq:4}   
\end{eqnarray}
where $H_z$, $M_{1z}$, and $M_{2z}$ are the projections onto the $z$-axis.

The effective fields acting on the layer magnetizations, $H_1 = - \partial E / \partial M_{1z}$ and $H_2 = - \partial E / \partial M_{2z}$, are given by
\numparts
\begin{eqnarray}
H_1 = -\left( J_{\text{dir}} + J_{\text{indir}} \right) M_{2z} + KM_{1z} + H_z &\label{eq:sub1} , \\
H_2 = -\left( J_{\text{dir}} + J_{\text{indir}} \right) M_{1z} + KM_{2z} + H_z &\label{eq:sub2} .
\end{eqnarray}
\endnumparts

Reorientation of the magnetization vectors occurs at points of the loss of equilibrium, when $H_1=0$, $H_2=0$, or both. From Eqs.~(\ref{eq:sub1}) and (\ref{eq:sub2}), the external field at which the magnetic states lose stability is obtained as
\numparts
\begin{eqnarray}
-\left( J_{\text{dir}} + J_{\text{indir}} \right) M_{2z} + KM_{1z} + H_z & = 0 \label{eq:sub3} , \\
-\left( J_{\text{dir}} + J_{\text{indir}} \right) M_{1z} + KM_{2z} + H_z & = 0 \label{eq:sub4} .
\end{eqnarray}
\endnumparts

The ferromagnetic-type state $M_{1z}=M_{2z}=\pm M_{\text{s}}$ loses stability at
\begin{equation}
     H_{\text{cr1}} = KM_{\text{s}} - \left( J_{\text{dir}} + J_{\text{indir}} \right) M_{\text{s}},
    \label{eq:7}
\end{equation}
while the antiferromagnetic state $M_{1z}=-M_{2z}= \pm M_{\text{s}}$ loses stability at
\begin{equation}
     H_{\text{cr2}} = KM_{\text{s}} + \left( J_{\text{dir}} + J_{\text{indir}} \right) M_{\text{s}}.
    \label{eq:8}
\end{equation}

From equations~(\ref{eq:7}) and (\ref{eq:8}), we find that the magnetic anisotropy field $H_{\text{a}} = K M_{\text{s}} $ and the interlayer exchange field $H_{\text{ex}} = \left( J_{\text{dir}} + J_{\text{indir}} \right) M_{\text{s}}$ can be expressed in terms of the critical fields as follows: $H_{\text{a}} = \left( H_{\text{cr1}} + H_{\text{cr2}}\right) /2 $.

This allows us to determine the exchange fields from experimental data.
The theoretically determined critical fields $H_{\text{cr1}}$ and $H_{\text{cr2}}$ should coincide with the experimentally measured fields $H_1$ and $H_2$, corresponding to the loss of stability of the film’s magnetic states. The temperature dependence of these fields is shown in Fig.~\ref{figure4}.

The temperature dependence of $H_{\text{a}}(T)$ and $H_{\text{ex}}(T)$ is presented in Fig.~\ref{figure4}~(b). They are obtained using $H_{\text{a}} (T) = \left[ H_1 (T) + H_2 (T) \right] / 2 $, $ H_{\text{ex}} (T) = \left[ H_1 (T) - H_2 (T) \right] / 2 $. 

As expected, $H_{\text{a}}(T)$ shows only a weak temperature dependence, whereas $H_{\text{ex}}(T)$ varies strongly with $T$.

Since the total interlayer exchange field is the sum of the direct and indirect contributions, $H_{\text{ex}} = H_{\text{ex}}^{\text{dir}} + H_{\text{ex}}^{\text{indir}}$ with $H_{\text{ex}}^{\text{dir}} = J_{\text{dir}} M_{\text{s}}$ and $H_{\text{ex}}^{\text{indir}} = J_{\text{indir}} M_{\text{s}}$, the observed behavior can be attributed to the temperature dependence of $J_{\text{indir}}(T)$. At high temperatures, direct exchange dominates, whereas the indirect contribution vanishes. Assuming $J_{\text{dir}}$ is temperature independent (valid at $T \ll T_{\text{c}}$), the strong $T$ dependence of $H_{\text{ex}}$ reflects the variation of $J_{\text{indir}}(T)$.

Experimental plots of $H_{\text{ex}}^{\text{dir}}(T)$ and $|H_{\text{ex}}^{\text{indir}}(T)|$ are shown in Fig.~\ref{figure4}~(c). The indirect exchange follows the temperature dependence of the paramagnetic susceptibility: it increases on cooling and tends to zero at high temperature.

\section{Conclusions}

Experimental results for the Fe(2)/Cr(0.4)/Fe$_{17.5}$Cr$_{82.5}$(0.9)/Cr(0.4)/Fe(2) five-layer structure demonstrate that the competition between direct and indirect RKKY-exchange interactions leads to a transition from antiferromagnetic alignment at high temperatures to ferromagnetic alignment at low temperatures. This behavior corresponds to a synthetic antiferromagnet with an effective, temperature-dependent interlayer exchange.

Our theoretical analysis shows that a thin paramagnetic layer incorporated into a synthetic RKKY antiferromagnet results in a temperature-dependent ferromagnetic contribution to the interlayer exchange, which modifies the overall magnetization behavior of the structure.

The magnetization reversal resembles that of classical metamagnets with a spin-flip transition, but in the above studied system the transition occurs in comparatively low magnetic fields. Although no intrinsic in-plane magnetocrystalline anisotropy was detected, the hysteresis loops indicate a field-induced uniaxial anisotropy with the easy axis aligned parallel to the applied field.

Our results demonstrate and explain the nature of thermally controlled AFM-to-FM switching in RKKY-based synthetic antiferromagnets, which adds additional functionality and can be useful in designing novel thermomagnetic devices.

\section*{Data availability}

All data supporting the findings of this study are available from the corresponding author upon reasonable request.

\ack

V.~Kalita and S.~Reshetniak acknowledge support from the IEEE Magnetics Society Program ``Magnetism for Ukraine 2024/2025'' (STCU Project No.~9918). A.~Kravets and V.~Korenivski acknowledge support from the Olle Engkvist Foundation (Grant No.~2020-207-0460) and the Swedish Strategic Research Council (SSF Grant No.~UKR24-0002).

\section*{References}

\bibliographystyle{iopart-num}
\bibliography{references}

\end{document}